\documentclass[a4paper]{article}\usepackage[]{graphicx}\usepackage[]{color}
\makeatletter
\def\maxwidth{ %
  \ifdim\Gin@nat@width>\linewidth
    \linewidth
  \else
    \Gin@nat@width
  \fi
}
\makeatother

\definecolor{fgcolor}{rgb}{0.345, 0.345, 0.345}

\usepackage{framed}
\makeatletter
 {\par\unskip\endMakeFramed%
 \at@end@of@kframe}
\makeatother

\definecolor{shadecolor}{rgb}{.97, .97, .97}
\definecolor{messagecolor}{rgb}{0, 0, 0}
\definecolor{warningcolor}{rgb}{1, 0, 1}
\definecolor{errorcolor}{rgb}{1, 0, 0}
\newenvironment{knitrout}{}{} 

\usepackage{alltt}
\usepackage[]{graphicx}
\usepackage[]{color}
\usepackage[left = 38mm, right = 25mm, top = 38mm, bottom = 25mm, headheight = 15pt]{geometry} 
\usepackage[comma]{natbib}
\usepackage{hyperref} 
\usepackage{setspace} 
\usepackage[]{mathalfa} \usepackage[]{amsfonts} \usepackage[]{amsmath} \usepackage{dsfont} \usepackage{amssymb} 
\usepackage{verbatim} 
\usepackage{authblk} 
\usepackage[final]{changes} 
\usepackage{relsize} 
\usepackage[singlelinecheck=false,justification=centering]{caption} 
\usepackage{subcaption} 
\usepackage{url} 
\usepackage{tikz} 
\allowdisplaybreaks 

\title{
  \textbf{A Network Epidemic Model for Online Community Commissioning Data} \\
}
\author[1,2]{Clement Lee}
\author[2]{Andrew Garbett}
\author[1]{Darren J Wilkinson}
\affil[1]{School of Mathematics and Statistics, Newcastle University}
\affil[2]{Open Lab, Newcastle University}
\IfFileExists{upquote.sty}{\usepackage{upquote}}{}
\begin{document}
\maketitle

\begin{abstract}
A statistical model assuming a preferential attachment network, which is generated by adding nodes sequentially according to a few simple rules, usually describes real-life networks better than a model assuming, for example, a Bernoulli random graph, in which any two nodes have the same probability of being connected, does. Therefore, to study the propogation of ``infection'' across a social network, we propose a network epidemic model by combining a stochastic epidemic model and a preferential attachment model. A simulation study based on the subsequent Markov Chain Monte Carlo algorithm reveals an identifiability issue with the model parameters. Finally, the network epidemic model is applied to a set of online commissioning data.
\end{abstract}

\section{Introduction} \label{sect.intro}
Social network analysis has been a popular research topic over the last couple of decades, thanks to the unprecedentedly large amount of internet data available, and the increasing power of computers to deal with such data, which details ties between people or objects all over the world. A lot of models have been developed to characterise and/or generate networks in various ways. One well-known class of models in the statistical literature is the exponential random graph model (ERGM), in which the probability mass function on the graph space is proportional to the exponential of a linear combination of graph statistics; see, for example, \cite{snijders02}. The Bernoulli random graph (BRG), in which any two nodes have the same probability of being connected, independent of any other pair of nodes, is a special case of an ERGM. Although the choice of graph statistics allows an ERGM to encompass networks with different characteristics, in general the ERGMs do not describe real-life networks well; see, for example, \cite{snijders02} and \cite{hgh08}.

Instead of characterising a network by graph statistics, such as the total number of degrees, the configuration model considers the sequence of the individual degrees; see, for example, \cite{newman10}, Chapter 13. Each node is assigned a number of half-edges according to its degree, and the half-edges are paired at random to connect the nodes. Despite its simple rule of network generation, the configuration model may contain multiple edges or self-connecting nodes, which might not occur in real-life networks. Also, the whole network is not guaranteed to be connected. Moreover, even though the individual degrees may be flexibly modelled by a degree distribution, they are not completely independent as they have to sum to an even integer.

One prominent feature of social networks in real life is that they are scale-free, which means that the degree distribution follows a power law (approximately); see, for example, \cite*{ajb99, ajb00}, and \cite*{swm05}. The preferential attachment {(PA)} model by \cite{ba99} is one widely known model \citep[Chapter~14]{newman10} that generates such a network with a few parameters and a simple rule. Other models also exist that characterise either the degree distribution, for instance the small-world model by \cite{ws98}, or other aspects such as how clustered the nodes in the network are \citep{vpv02}.

While the majority of the network models focus on the topology of the network, some models are developed to describe the dynamics within the network, in particular how fast information spreads with respect to the structure of the network. As spreading rumours or computer viruses through connections in a social network is similar to spreading a disease through real life contacts to create an epidemic, most of these models incorporate certain compartment models in epidemiology. For instance, the Susceptible-Infectious-Recovered (SIR) model splits the population into three compartments according to the stage of the disease of each individual. A susceptible individual becomes infectious upon contact with an infectious individual, and recovers after a random period. Traditionally, the infectious period and the contacts made by an infected individual are assumed to follow an exponential disribution and a homogeneous Poisson Process, respectively. While these assumptions may be unrealistic for real life data, they are useful as the epidemic process is now Markovian. The dynamics of compartment sizes over time can usually be characterised by a small number of parameters in the rate matrix, which is used to obtain the transition probabilities through the Kolmogorov's equations; see, for example, \cite{wilkinson11}, Section 5.4. While other kinds of compartment models can be formulated in a similar way, some models depart from the Markovian assumptions, and will be discussed later. For more details on the SIR model and its variants, see, for example, \cite{ab00}.

Often implicitly assumed in such compartment models is that the epidemic is homogeneous mixing, that is, each individual can interact uniformly with all other individuals in the community he/she belongs to. However, this is not the case when it comes to network epidemics, as one can only infect and be infected by their neighbours in the network, and the collection of neighbours differs from individual to individual. Therefore, modelling an epidemic on a structured population requires relaxing the homogeneous mixing assumption. Instead of assuming the same set of values for the parameters governing the dynamics, one approach is to apply a separate set of parameter values to, for example, each individual or all individuals with the same degree. Such an approach focuses on the modelling side, and is dominant in the physics literature. A comprehensive review is provided by \cite*{pcvv15}.

Our work on network epidemic modelling is motivated by a data set from App Movement\footnote{\url{https://app-movement.com}}, which is an online platform that enables communities to propose and design community-commissioned mobile applications \citep*{gcjo16}. The process of generating the application starts with a community creating a campaign page and sharing it via online social networks. If we view an individual having seen a campaign and in turn promoting it as being ``infected'' (and ``infectious'' simultaneously), then the process of sharing a campaign can be compared to spreading a real-life virus to create an epidemic. The main difference is that such an infectious individual cannot potentially infect anyone in the population but only those connected to them on the social networks. For one campaign, the cumulative count of infected and the network of infected users are plotted in Figures \ref{fig.eda_I} and \ref{fig.eda_P}, respectively. The former deviates from the typical S shape of a homogeneous mixing epidemic, while the latter displays star-like structures and long paths, which typical features in real-life networks. It should be noted that this does not represent the complete underlying network $\mathcal{G}$, which is usually unknown. 

\begin{knitrout}
\definecolor{shadecolor}{rgb}{0.969, 0.969, 0.969}\color{fgcolor}\begin{figure}[htb!]

{\centering \includegraphics[width=\maxwidth]{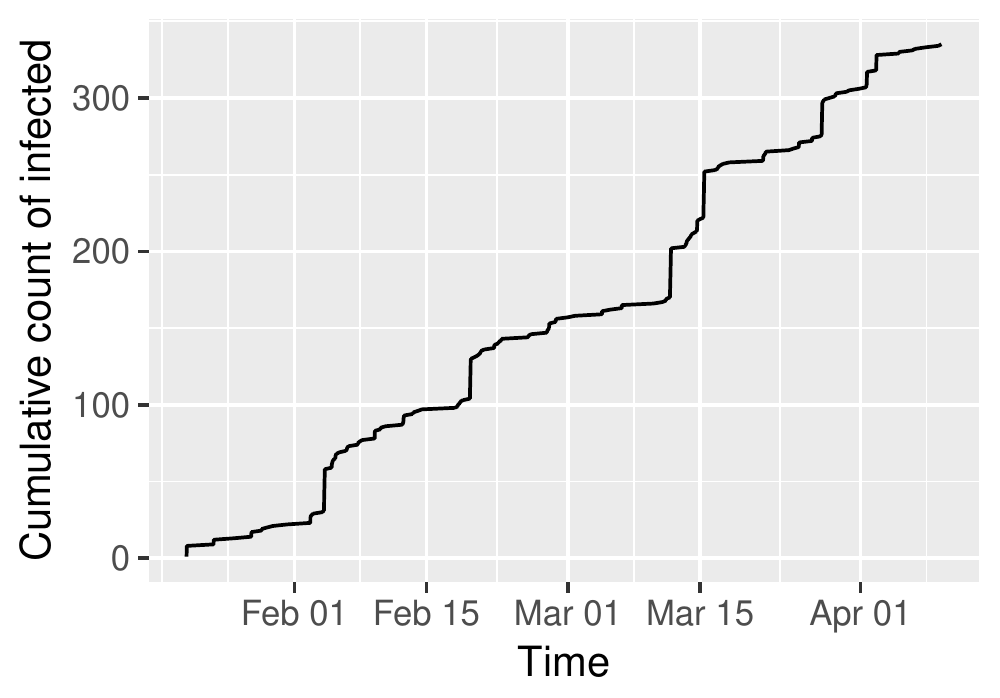} 

}

\caption[Cumulative count of infected for an epidemic of a campaign]{Cumulative count of infected for an epidemic of a campaign. The time points where the count increments are the infection times $\mathbf{I}$.}\label{fig.eda_I}
\end{figure}

\end{knitrout}
\begin{knitrout}
\definecolor{shadecolor}{rgb}{0.969, 0.969, 0.969}\color{fgcolor}\begin{figure}[htb!]

{\centering \includegraphics[width=\maxwidth]{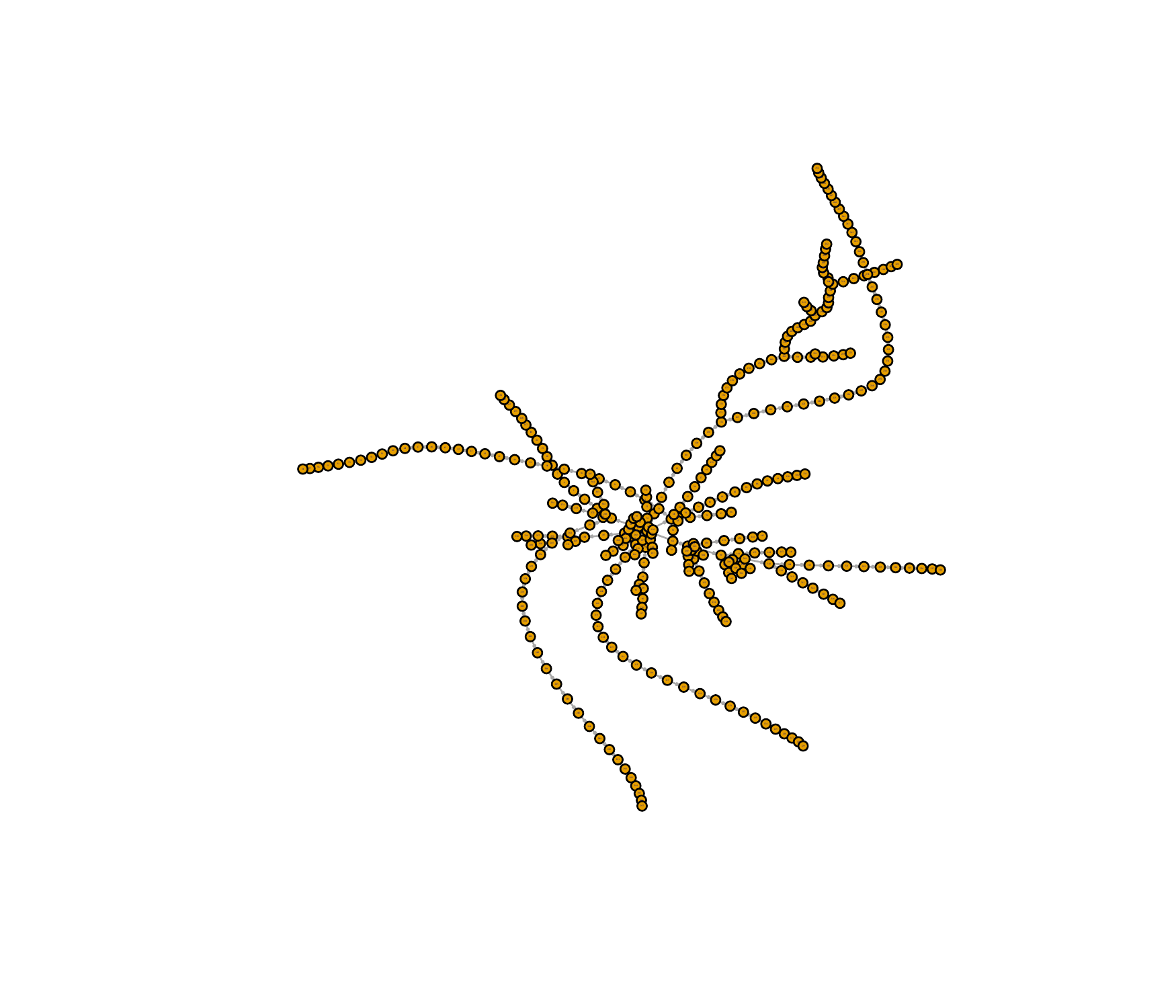} 

}

\caption{Network representation of the transmission tree $\mathcal{P}$ for the same campaign epidemic as shown in Figure \ref{fig.eda_I}.}\label{fig.eda_P}
\end{figure}

\end{knitrout}

Due to the difference in the data being applied to, as well as the inclination towards inference, epidemic models in the statistics literature provide a stark contrast from the classical compartment model, not only with respect to the network issue. First, to accommodate heterogeneities in mixing, \cite{bms97} and \cite{bko11} proposed models which incorporate two levels and three levels of mixing, respectively. Each individual belongs to both the global level and one or more local levels, such as household, school or workplace, and homogeneous mixing is assumed to take place at each level but with a separate rate. Such models are prompted by data with detailed information of these local level structures each individual belongs to, such as the 1861 Hagelloch measles outbreak data analysed by \cite{bko11}. Second, some SIR models and their variants relax the assumption that the infectious period follows the exponential distribution, essentially rendering the epidemic process non-Markovian. For instance, \cite{sg02} used the Weibull distribution, while \cite{nr05} and \cite*{gwh12} used the Gamma distribution. In general, the compartment dynamics cannot be represented by a simple differential equation. Third, information is often missing in epidemic datasets, such as the infection times and, if a network structure is assumed, the actual network itself. Therefore, models are developed with a view to inferring these missing data, usually achieved by Markov Chain Monte Carlo (MCMC) algorithms. Examples of models which impose a network structure include {\cite{bo02}}, \cite{nr05}, \cite{rm08} and \cite*{gwh11}. In the data considered by these authors, no covariates exist to inform if two individuals are neighbours in the network, and the edge inclusion probability parameter is assumed to be the same for any two individuals in the network. Essentially the underlying network is a BRG, which yields a Binomial (or approximately Poisson) degree distribution. Such a network model seems unrealistic for our App Movement data, compared to a model that generates a scale-free network or utilises a power law type degree distribution.

In view of the differences in objectives and applications shown above, we propose a network epidemic model as an attempt to narrow the gap in the literature. We focus on a Susceptible-Infectious (SI) model, in which the epidemic process takes place on a network which is assumed to be built from the PA model, thus deviating from a BRG. When it comes to inference, the data contains the infection times and potentially the transmission tree, while the underlying network is unknown and therefore treated as latent variables. We aim at simultaneously inferring the infection rate parameter, the parameters governing the degree distribution, and the latent structure of the network, in terms of the posterior edge inclusion probabilities, by using an MCMC algorithm. While the choice of the SI model is due to the data in hand, we believe the model structure and algorithm introduced can be extended to other compartment models.

The rest of the article is divided as follows. The latent network epidemic model is introduced in Section \ref{sect.model}. Its likelihood and its associated MCMC algorithm are derived in Section \ref{sect.lik_inf}. They are then applied to two sets of simulated data in Section \ref{sect.sim}, and a set of real online commissioning data in Section \ref{sect.app}. Section \ref{sect.con} concludes the article.

\section{Model} \label{sect.model}
In this section we introduce the latent network SI epidemic model. Describing the formation of the network and the epidemic separately will facilitate the derivation of the likelihood in the next section. The notations and definitions are kept to be similar to those in \cite{bo02} and \cite{gwh11}.

Consider an epidemic in a closed population of size $m$. Let $\mathbf{I} = (I_1, I_2, \ldots, I_m)$ denote the ordered vector of infection times, where $I_i$ is the infection time of individual $i$, and $I_i \leq I_j$ for any $i<j$. We assume that the first individual is the only initial infected individual. In order to have a temporal point of reference, only the times of $m-1$ infections will be random, and so we define $\tilde{\mathbf{I}}=\mathbf{I}-I_1=(\tilde{I}_1=0,\tilde{I}_2,\ldots,\tilde{I}_m)$ for convenience. We also assume that the observation period is long enough to include all infections.

Next, consider the undirected random graph $\mathcal{G}$ of $m$ nodes which represents the social structure of the population, in which the node $i$ represents the $i^{\text{th}}$ individual. Using the adjacency matrix representation, if individuals $i$ and $j$ are socially connected, we write $\mathcal{G}_{ij}=1$ and call them neighbours of each other, $\mathcal{G}_{ij}=0$ otherwise. In this sense $\mathcal{G}_{ij}$ can be interpreted as a potential edge of $i$ and $j$. We also assume symmetry in social connections and that each individual is not self-connected, that is, $\mathcal{G}_{ij} = \mathcal{G}_{ji}$ and $\mathcal{G}_{ii}=0$, respectively, for $1\leq i,j\leq m$.

To characterise $\mathcal{G}$, we use a modified version of the PA model by \cite{ba99}, which generates a network by sequentially adding nodes into it. This requires an order of how the nodes enter the network, which is not necessarily the same as the epidemic order. Therefore, we define a vector random variable of the network order, denoted by $\boldsymbol{\sigma}=(\sigma_1,\sigma_2,\ldots,\sigma_m)$, whose support is all $m!$ possible permutations of $\{1,2,\ldots,m\}$. Node $\sigma_i~(1\leq i\leq m)$, labelled by the epidemic order, is the $i^{\text{th}}$ node that enters the network. Such order is mainly for the sake of characterisation using the PA model, and the network is assumed to have formed before the epidemic takes place, and remain unchanged throughout the course of the epidemic. Such an assumption is reasonable because the timescale of an epidemic is usually much smaller than that of network formation, the process of which is described next.

\subsection{Sequence of new edges} \label{sect.model_edges}

Initially, there are two nodes $\sigma_1$ and $\sigma_2$ which are connected i.e. $\mathcal{G}_{\sigma_1\sigma_2}=1$. When node $\sigma_i~(3\leq i\leq m)$ enters the network, it connects to $X_i$ existing nodes, where $X_i$ follows a censored Poisson distribution with parameter $\mu$ and support $\{1,2,\ldots,i-1\}$, that is,
\begin{align}
  \Pr(X_i=x)=\left\{\begin{array}{ll}
  e^{-\mu}(1+\mu), & x=1, \\
  \displaystyle\frac{e^{-\mu}\mu^x}{x!}, & x=2,3\ldots,i-2, \\
  \displaystyle\sum_{z=i-1}^\infty\frac{e^{-\mu}\mu^z}{z!}, & x=i-1. \end{array} \right.\label{eqn.edges}
\end{align}
Independence is assumed between $X_i$ and $X_j$ if $i\neq j$. We model the number of new edges as a random variable because using a constant number of new edges, denoted by $\mu_0$, which is what the original model by {\cite{ba99}} did, fixes the total number of edges to $(m-2)\mu_0+1$ and makes the model too restrictive. Empirically this censored distribution performs better than a \textit{truncated} Poission distribution in terms of identifying $\mu$.

\subsection{Attaching edges to nodes} \label{sect.model_network}

When node $\sigma_i$ joins the network, according to the original PA rule, an existing node $\sigma_j~(1\leq j<i)$ gets connected to node $\sigma_i$, that is, $\mathcal{G}_{\sigma_i\sigma_j}=1$, with probability proportional to its current degree $\sum_{k=1}^{i-1}\mathcal{G}_{\sigma_k\sigma_j}$. To allow the degree of PA to vary, we allow such probability to be a mixture of the current degree and how recently the node has joined the network. To be more specific, the process of choosing $x_i$ nodes is equivalent to obtaining a weighted random sample \textit{without} replacement from $\{1,2,\ldots,i-1\}$, with the weight assigned to node $\sigma_j$ equal to $w_j$, where
\begin{align}
  w_j=(1-\gamma)\frac{\sum_{k=1}^{i-1}\mathcal{G}_{\sigma_k\sigma_j}}{~\sum_{l=1}^{i-1}\sum_{k=1}^{i-1}\mathcal{G}_{\sigma_k\sigma_l}~}+\gamma\frac{j}{~\sum_{l=1}^{i-1}l~}, \label{eqn.model_weights}
\end{align}
where $\gamma\in[0,1]$ and can be seen as the parameter governing the degree of PA. When $\gamma=0$, this reduces to the original PA rule. When $\gamma$ increases, more weights are given to latter nodes, and the inequality in the degrees of the nodes is reduced. Such inequality reduction is facilitated by assigning weights according to how recent the nodes join the network, rather than equal weights, in the non-PA component. Note that, however, even in the extreme case where $\gamma=1$, where the degree distribution is unimodal and closer to symmetry, the model does \textit{not} reduce to a BRG, where the degree distribution is Binomial with parameters $(m-1,p)$, where $p$ is the edge inclusion probability, but provides a crude approximation to it.

\subsection{Constructing the epidemic} \label{sect.model_epidemic}

The Markovian epidemic process is constructed as follows. At time $0$, the whole population is susceptible except individual 1, who is infected. Once infected at time $\tilde{I}_i$, individual $i$ makes infectious contacts at points of a homogeneous Poisson process with rate $\beta\sum_{j=1}^m\mathcal{G}_{ij}$ with its neighbours (according to $\mathcal{G}$), and stays infected until the end of the observation period. The random transmission tree $\mathcal{P}$, with the same node as $\mathcal{G}$ and whose root is the node labelled 1, can be constructed simultaneously. If individual $i$ makes infectious contact at arbitrary $t_0$ (governed by the aforementioned Poisson process) with susceptible neighbour $j$, we write $\mathcal{P}_{ij}=1$, again using the adjacency matrix representation. This implies $\tilde{I}_j=t_0$, and $\mathcal{P}_{ji}=0$ as individual $i$ cannot be re-infected. Also, $\mathcal{P}_{ij}=1$ implicity implies that $\mathcal{G}_{ij}=(\mathcal{G}_{ji}=)1$, as the epidemic can only spread through social connections i.e. the edges in $\mathcal{G}$. Also, we assume $\mathcal{P}_{ii}=0$ as any individual cannot be infected by themselves.

\section{Likelihood and inference} \label{sect.lik_inf}
We proceed to compute the likelihood, denoted by $L$, as a function of $\beta$, $\mu$, $\gamma$ and $\boldsymbol{\sigma}$. We assume both $\mathcal{G}$ and $\mathcal{P}$ are given because, as argued by \cite{bo02} and \cite{gwh11}, it is easier to condition on $\mathcal{G}$ and $\mathcal{P}$ in order to calculate $L$, and, if they are unobserved, include them as latent variables in the inference procedure. Two conditional independence assumptions need to be noted. Because of the Markovian nature of the epidemic, $\mathcal{P}$ and $\tilde{\mathbf{I}}$ are independent given $\mathcal{G}$. It is also common that the data $(\{\tilde{\mathbf{I}},\mathcal{P}\})$ and (a subset of) the parameters $(\mu,\gamma,\boldsymbol{\sigma})$ are independent \textit{apriori}, given $\mathcal{G}$, when models are formulated by centred parameterisations \citep*{prs03}. Therefore, the likelihood can be broken down into the following components:
\begin{align}
  L:=L(\beta,\mu,\gamma,\boldsymbol{\sigma})
  &=\pi(\mathcal{P},\tilde{\mathbf{I}},\mathcal{G}|\beta,\mu,\gamma,\boldsymbol{\sigma})\nonumber\\
  &=\pi(\mathcal{P},\tilde{\mathbf{I}}|\mathcal{G},\beta,\mu,\gamma,\boldsymbol{\sigma})\times
  \pi(\mathcal{G}|\beta,\mu,\gamma,\boldsymbol{\sigma})\nonumber\\
  &=\pi(\mathcal{P}|\mathcal{G})\times\pi(\tilde{\mathbf{I}}|\mathcal{G},\beta)\times
  \pi(\mathcal{G}|\mu,\gamma,\boldsymbol{\sigma}).\label{eqn.lik}
\end{align}
The dropping of any unrelated quantities can be explained by how the network and the epidemic are constructed in Section \ref{sect.model}, and is demonstrated in the derivations of each component in Appendix \ref{sect.appendix_lik}, the results of which are given below:
\begin{align}
  \pi(\mathcal{P}|\mathcal{G})
  =&~\mathlarger\prod_{j=2}^{m}\mathbf{1}\left\{\sum_{i=1}^{j-1}\mathcal{P}_{ij}=1\right\}\left(\sum_{i=1}^{j-1}\mathcal{G}_{ij}\right)^{-1}
  \mathlarger\prod_{1\leq i<j\leq m}(1-\mathcal{P}_{ji})\mathbf{1}\left\{\mathcal{P}_{ij}\leq\mathcal{G}_{ij}\right\}, \label{eqn.lik_p} \\
  \pi(\tilde{\mathbf{I}}|\mathcal{G},\beta)
  =&~\beta^{m-1}\exp\left(-\beta\mathlarger\sum_{1\leq i<j\leq m}\mathcal{G}_{ij}(\tilde{I}_j-\tilde{I}_i)\right), \label{eqn.lik_i} \\
  \pi(\mathcal{G}|\mu,\gamma,\boldsymbol{\sigma})
  =&~\mathlarger{
    e^{-(m-2)\mu}\mu^{|\mathcal{G}|-1}
    (1+\mu)^{\sum_{i=3}^{m}\boldsymbol{1}\{\sum_{j=1}^{i-1}\mathcal{G}_{\sigma_i\sigma_j}=1\}}
  } \nonumber \\*
  &\times  \mathlarger\prod_{i=3}^m \left[\sum_{z=i-1}^{\infty}\frac{\mu^z}{z!}\left/\frac{\mu^{i-1}}{(i-1)\mathlarger!}\right.\right]^{\boldsymbol{1}\{\sum_{j=1}^{i-1}\mathcal{G}_{\sigma_i\sigma_j}=i-1\}} \nonumber \\*
  &\times  \mathlarger\prod_{i=3}^m\left[w_1^{\mathcal{G}_{\sigma_i\sigma_1}}\mathlarger\prod_{j=2}^{i-1}\left(\frac{w_j}{1-\sum_{k=1}^{j-1}w_k}\right)^{\mathcal{G}_{\sigma_i\sigma_j}}\right].\label{eqn.lik_L1L2}
\end{align}

We can proceed to inference because the likelihood \eqref{eqn.lik} can be expressed explicitly as the product of \eqref{eqn.lik_p}-\eqref{eqn.lik_L1L2}. However, this complete likelihood is only useful for inference when $\mathcal{G}$ (and $\mathcal{P}$) is given or known, which is usually not the case in real-life applications. As each of the $\binom m 2$ potential edges is a binary random variable, integrating $\mathcal{G}$ out does not seem feasible as we will have to average over all $2^{\binom m 2}$ possibilities. Also, unlike the scalar parameters $(\beta,\mu,\gamma)$, the support of $\boldsymbol{\sigma}$ is the permutation space of $\{1,2,\ldots,m\}$. It is not meaningful to calculate a certain kind of point estimate of $\boldsymbol{\sigma}$ and quantify its uncertainty using a frequentist approach. It is therefore quite natural to consider Bayesian inference, in which $\mathcal{G}$ is considered as latent variables, the posterior probabilities of which are to be computed simultaneously with those of the model parameters. It is sensible to assume the infection rate $\beta$, which relates to the intrinsic properties of the disease, to be independent of $\mu$ and $\gamma$ \textit{apriori}, which relate to the properties of the network. We assign the following independent and vaguely informative priors:
\begin{align}
  \beta&\sim\text{Gamma}(a_{\beta}=1,b_{\beta}=0.001),\nonumber\\
  \mu&\sim\text{Gamma}(a_{\mu}=1,b_{\mu}=0.001),\nonumber\\
  \gamma&\sim U[0,1],\nonumber\\
  \pi(\boldsymbol{\sigma})&=(m!)^{-1}\boldsymbol{1}\{\boldsymbol{\sigma}\text{ is a permutation of }\{1,2,\ldots,m\}\},\label{eqn.inf_prior}
\end{align}
where $a/b$ is the mean of a random variable $X\sim\text{Gamma}(a,b)$. By Bayes' theorem, we have
\begin{align}
  \pi(\mathcal{G},\beta,\mu,\gamma,\boldsymbol{\sigma}|\mathcal{P},\tilde{\mathbf{I}})
  \propto\pi&(\mathcal{P},\tilde{\mathbf{I}},\mathcal{G},\beta,\mu,\gamma,\boldsymbol{\sigma})\label{eqn.inf_joint}\\
  =\pi&(\mathcal{P},\tilde{\mathbf{I}},\mathcal{G}|\beta,\mu,\gamma,\boldsymbol{\sigma})
  \times\pi(\beta,\mu,\gamma,\boldsymbol{\sigma})\nonumber\\
  =\pi&(\mathcal{P}|\mathcal{G})\times\pi(\tilde{\mathbf{I}}|\mathcal{G},\beta)\times\pi(\mathcal{G}|\mu,\gamma,\boldsymbol{\sigma})\times\pi(\beta)\pi(\mu)\pi(\gamma)\pi(\boldsymbol{\sigma}) \label{eqn.inf_post}
\end{align}
As the posterior density, up to a proportionality constant, can be obtained explicitly as the product of \eqref{eqn.lik_p}-\eqref{eqn.inf_prior}, a natural candidate for inference is MCMC. We use a component-wise Metropolis-within-Gibbs (MWG) algorithm, described in detail in Appendix \ref{sect.appendix_mcmc}, in which each of the parameters $(\beta, \mu, \gamma, \boldsymbol{\sigma})$ is sampled conditional on the other three parameters and the whole of $\mathcal{G}$, while each potential edge of $\mathcal{G}$ is sampled conditional on all parameters and other potential edges of $\mathcal{G}$.

\section{Simulation study} \label{sect.sim}

A simulation study is carried out to examine if the inference algorithm in Appendix \ref{sect.appendix_mcmc} can recover the true values of the parameters used to simulate from the model in Section \ref{sect.model}. Specifically, we set $m=$ 70 and consider all combinations of the following true values: $\gamma=$ 0,0.2,0.5,0.8,1, $\beta=$ 0.4, and $\mu=$ 4,6,8,10. For each of the 20 combinations, we first simulate the PA network, then simulate the epidemic on the network. Because of how we construct and simulate from the model, we have complete information on the underlying graph $\mathcal{G}$, the transmission tree $\mathcal{P}$, and the infection times $\tilde{\mathbf{I}}$. When $\mathcal{G}$ is given together with $\mathcal{P}$ and $\tilde{\mathbf{I}}$, the MCMC algorithm only needs to be applied to $\beta$, $\mu$, $\gamma$ and $\boldsymbol{\sigma}$, and it successfully recovers each of the three scalar parameters. Also, the posterior correlations between $\beta$ and $\mu$ and between $\beta$ and $\gamma$ are both close to zero, which makes sense because of the independence conditional on $\mathcal{G}$, according to \eqref{eqn.inf_post}. However, we should focus on how good the algorithm is at inferring $\mathcal{G}$ given $\mathcal{P}$ and $\tilde{\mathbf{I}}$ only, because $\mathcal{G}$ is usually unknown in real-life data, while $\mathcal{P}$ being known is motivated by the data set in Section \ref{sect.app}. Therefore, the complete MCMC algorithm for $\beta$, $\mu$, $\gamma$, $\boldsymbol{\sigma}$ and $\mathcal{G}$ is applied to the same set of simulated data for each parameter combination.

A chain of 20000 iterations (no thinning) is obtained, with the first 10000 iterations discarded as burn-in. For $\mu$ and $\gamma$, a random walk Metropolis (RWM) step with a Gaussian proposal is used, and this step is modified into an adaptive one during burn-in in order to tune the proposal standard deviation, using the method outlined by, for example, {\cite{xn14}, Section 3}. The algorithm in general fails to correctly identify any of the three scalar parameters. The true value of $\gamma$ is plotted against its posterior distribution at different true values of $\mu$ in Figure \ref{fig.sim_1st_gamma}. The absence of significant linear correlation between the posterior mean of $\gamma$ and its true value, and the huge degree of uncertainty at all true values considered suggest that it is difficult to identify $\gamma$ in particular. Similar to Figure \ref{fig.sim_1st_gamma}, for different true values of $\mu$ we also plot the true value of $\mu$ against its posterior distribution in Figure \ref{fig.sim_1st_mu}, which suggests that, intriguingly, the latter does not depend on the former. This is still the case if we further fix $\gamma$ to its true value.

\begin{knitrout}
\definecolor{shadecolor}{rgb}{0.969, 0.969, 0.969}\color{fgcolor}\begin{figure}
\includegraphics[width=\maxwidth]{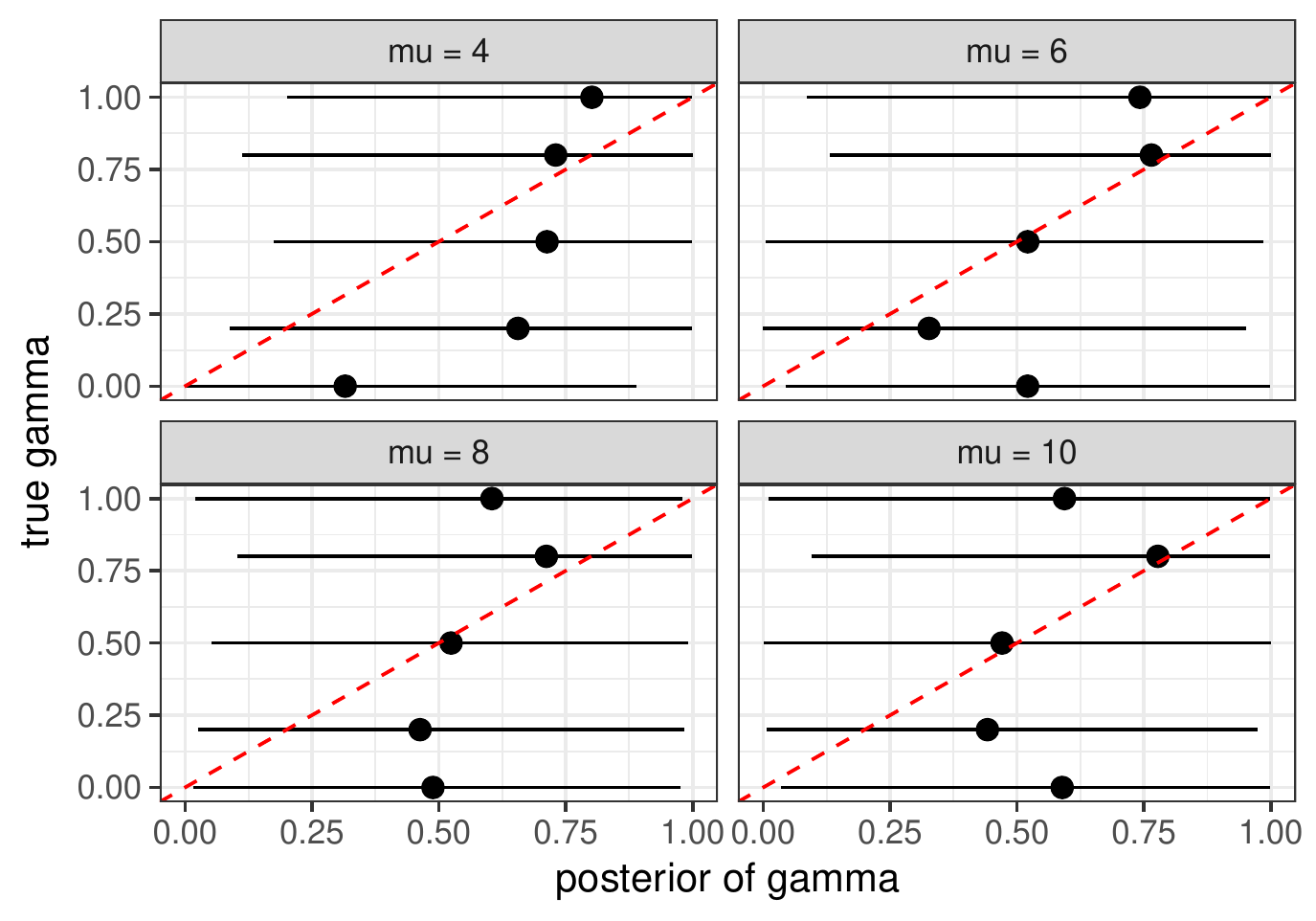} \caption[True value of $\gamma$ against its posterior distribution (horizontal line) and mean (dot), at different true values of $\mu$]{True value of $\gamma$ against its posterior distribution (horizontal line) and mean (dot), at different true values of $\mu$. The dashed line is the line $y=x$.}\label{fig.sim_1st_gamma}
\end{figure}

\end{knitrout}

\begin{knitrout}
\definecolor{shadecolor}{rgb}{0.969, 0.969, 0.969}\color{fgcolor}\begin{figure}
\includegraphics[width=\maxwidth]{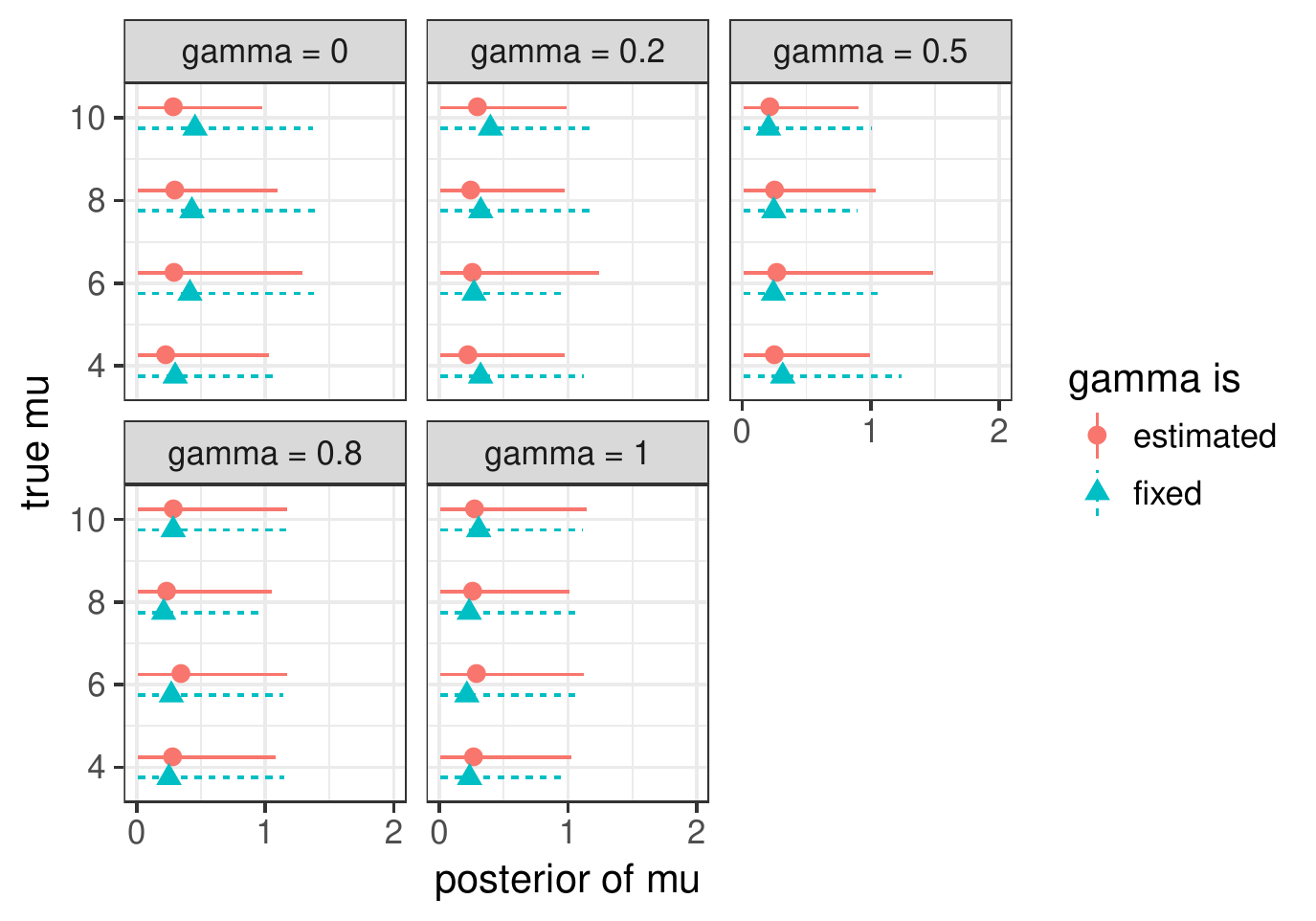} \caption[True value of $\mu$, slightly adjusted for visibility, against its posterior distribution (horizontal line) and mean (symbol), at different true values of $\gamma$, which is estimated (circle) or fixed (triangle) to its true value in the MCMC algorithm]{True value of $\mu$, slightly adjusted for visibility, against its posterior distribution (horizontal line) and mean (symbol), at different true values of $\gamma$, which is estimated (circle) or fixed (triangle) to its true value in the MCMC algorithm.}\label{fig.sim_1st_mu}
\end{figure}

\end{knitrout}

The identifiability issue prompts us to consider dropping or fixing at least one of $\beta$, $\mu$ and $\gamma$. While $\beta$ and $\mu$ are essential to characterise the epidemic and the network, respectively, leaving $\gamma$ out means we do not allow the degree of how preferentially attaching the network is to vary. Therefore $\gamma$ is fixed to be 0, which is equivalent to the network model being reduced to the original PA model, and will not be estimated.

A second simulation study is carried out, this time with $m$ being allowed to take different values, namely $m=$ 30,50,70, combined with the following true values: $\beta=$ 0.4 and $\mu=$ 4,6,8,10. For each parameter combination, we also allow different proportions of $\mathcal{G}$ in the simulated data to be known in addition to $\mathcal{P}$. A proportion of 0 means only $\mathcal{P}$ is known, while a proportion of 1 means both $\mathcal{P}$ and $\mathcal{G}$ are given. The true value of $\mu$ against its posterior distribution is plotted for different combinations of $m$ and proportions of $\mathcal{G}$ in Figure \ref{fig.sim_G_percent_mu}. The posterior of $\mu$ again shows no correlation with its true value in the first row, which corresponds to no $\mathcal{G}$ given at all, but it converges towards its true value as the proportion goes to 1. Also, it is now possible to recover the true value of $\mu$, with even, say, a quarter of the potential edges of $\mathcal{G}$ additional to $\mathcal{P}$.

\begin{knitrout}
\definecolor{shadecolor}{rgb}{0.969, 0.969, 0.969}\color{fgcolor}\begin{figure}
\includegraphics[width=1\linewidth,height=8.5in]{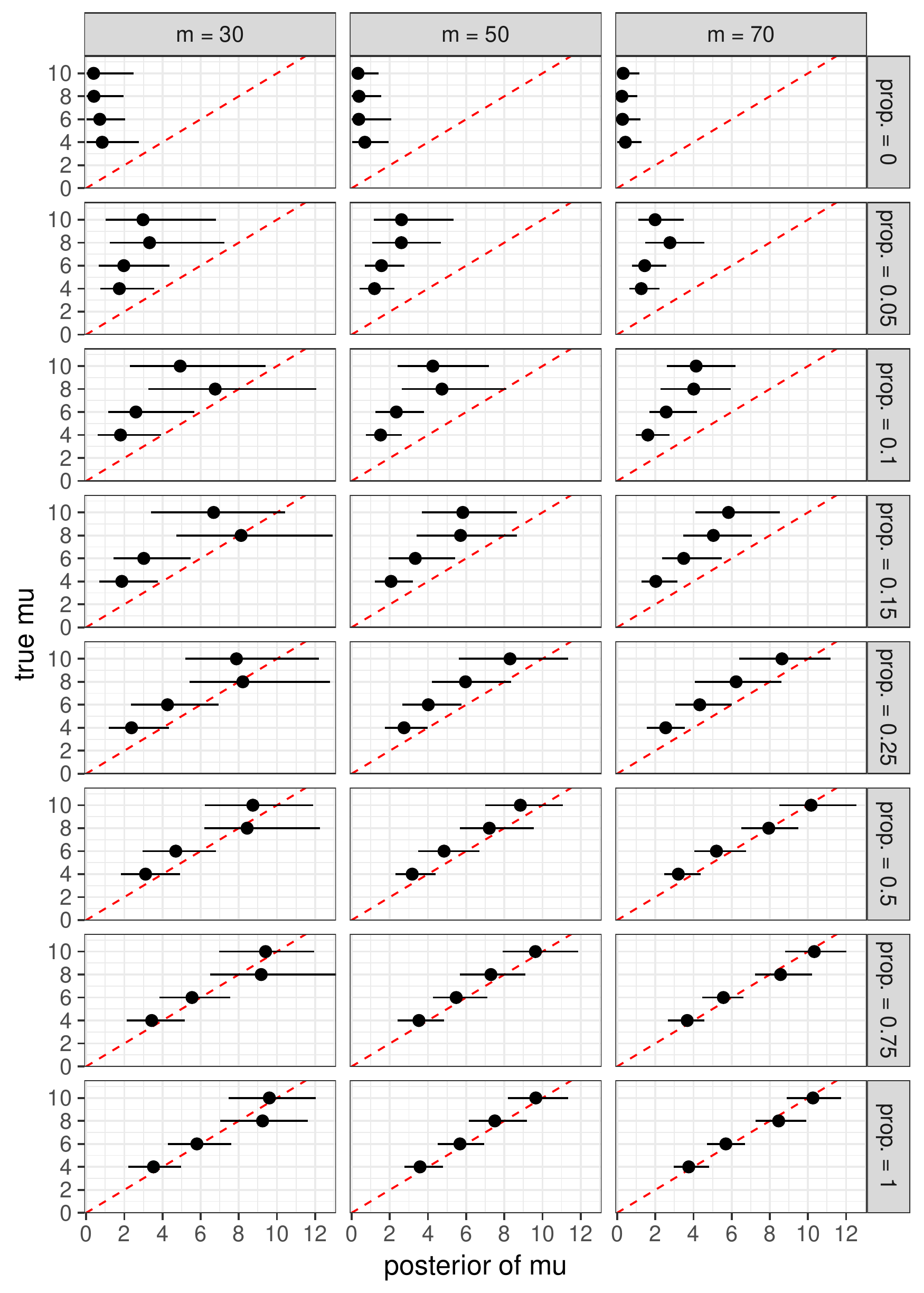} \caption[True value of $\mu$ against its posterior distribution (horizontal line) and mean (dot), at different combinations of $m$ and proportions of $\mathcal{G}$ known additional to those implied by $\mathcal{P}$]{True value of $\mu$ against its posterior distribution (horizontal line) and mean (dot), at different combinations of $m$ and proportions of $\mathcal{G}$ known additional to those implied by $\mathcal{P}$. The dashed line is the line $y=x$.}\label{fig.sim_G_percent_mu}
\end{figure}

\end{knitrout}

Rather than looking at the identifiability of one parameter alone, we can investigate the product $\alpha:=\beta\times\mu^{*}$, the posterior of which can be obtained post inference, where $\mu^{*}:=\mu+e^{-\mu}$. Plotting the true value of $\alpha$ against its posterior (not shown) in the similar way to Figure \ref{fig.sim_G_percent_mu} reveals that it is identifiable regardless of its true value, $m$, or the proportion of $\mathcal{G}$ given. The introduction of $\mu^{*}$ is due to that the mean of the distribution in \eqref{eqn.edges} is approximately $\mu+e^{-\mu}$ when $i$ is large. As $\alpha$ is the product of the (unscaled) epidemic rate and the average network connectedness, we can interpret it as the \textit{network scaled epidemic rate}. Epidemics on two different networks are comparable through this parameter if the two networks have similar values of $\mu^{*}$.

Such findings regarding $\alpha$ can be explained by looking the results of one parameter combination in detail. The joint posterior density in Figure \ref{fig.sim_G_percent_density2d} displays an inverse relationship between $\beta$ and $\mu^{*}$ when no $\mathcal{G}$ is given, echoing the findings by {\cite{bo02}}, who showed the inverse relationship between $\beta$ and the edge inclusion probability parameter $p$ in their BRG model, and argued that ``the model parameterisation permits different explanations of the same outcome''. This means that we cannot simultaneously identify the parameters that characterise the epidemic rate and the network connectedness, respectively, and being able to identify one relative to the other is as good as we can do.

\begin{knitrout}
\definecolor{shadecolor}{rgb}{0.969, 0.969, 0.969}\color{fgcolor}\begin{figure}
\includegraphics[width=1\linewidth,height=2.8in]{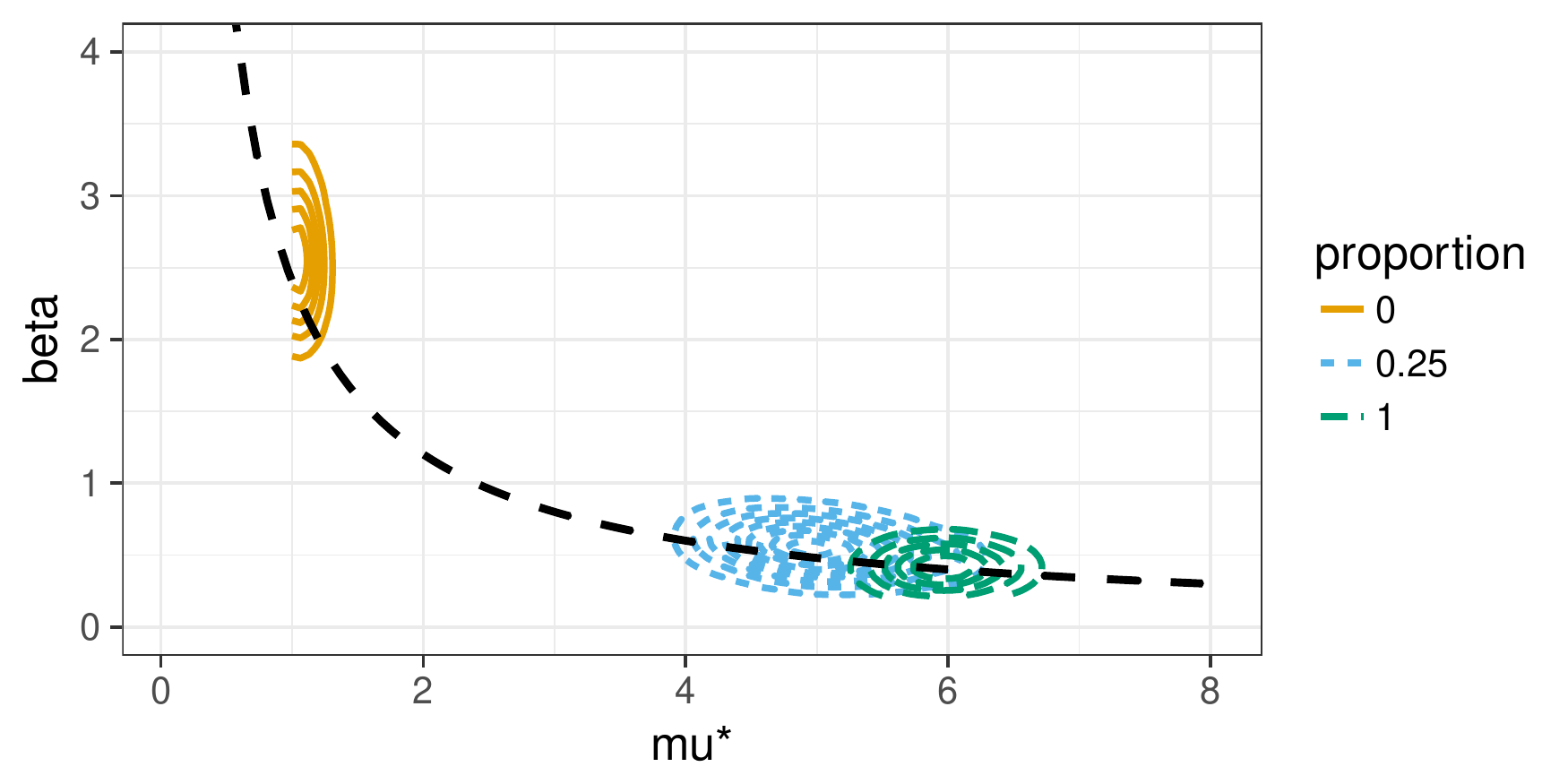} \caption{Joint posterior density plot of $\beta$ and $\mu^{*}$ at selected proportions of $\mathcal{G}$ given, for a simulated data set with $m=50$ and true values of the parameters $(\beta, \mu, \gamma)=(0.4, 6.0, 0.0)$. The black dashed curve is the line $xy=2.401$, the product of the true values of $\beta$ and $\mu^{*}$. As $\mu^{*}=\mu+e^{-\mu}$ is bounded below by 1, the contours where the proportion of $\mathcal{G}$ is 0 are truncated.}\label{fig.sim_G_percent_density2d}
\end{figure}

\end{knitrout}

\section{Application} \label{sect.app}
Before applying the proposed model to its data set introduced in Section \ref{sect.intro}, we shall describe App Movement in detail. This platform removes the resource constraints around mobile application development through providing an automated process of designing and developing mobile applciations. The process begins with the Support Phase whereby a community creates a campaign page in response to a community need and engages the community in supporting the concept through promoting and sharing the campaign on online social networks. When the target of 150 members supporting the campaign within 14 days has been met to ensure an active user base, the campaign proceeds to the Design Phase, in which ideas regarding the design of the mobile application are being voted on. Once supporters have cast their votes, the platform incorporates the highest rated design decisions and automatically generates the mobile application. Since its launch in February 2015, App Movement has been adopted by over 50,000 users supporting 111 campaigns, 20 of which have been successful in reaching their target number of supporters, with 18 generated mobile applications currently available in the Google Play Store and Apple App store, for iOS and Android devices, respectively.

The design of logging the usage of App Movement enables us to convert the data into a format suitable for modelling and fitting. To illustrate this, assume that user 1 shares a campaign page, with uniform resource locator (URL) A, on certain social network. When user 2, who is connected to user 1 and has never viewed the campaign page, clicks on URL A, a new URL B unique to user 2, directing to a page with the same contents of the campaign, is created. Any subsequent visits to the same page of users 1 and 2 will be redirected to the same URLs A and B, respectively. Therefore, within each campaign, there is a 1-1 relationship between the URLs and the users. We can say that user 1 infects user 2, at the time when URL B is created. Similarly, the users associated with the URLs created by clicking URL B can be said to be infected by user 2. This process is illustrated in the flow diagram in Figure \ref{fig.flow}. By carrying out this process of connecting users with those who infected them until we reach, in tree terminology, the root and all the leaves, we end up with both the transmission tree and the infection times of the epidemic of the campaign sharing. The inference outlined in Section \ref{sect.lik_inf} can then be carried out.

\usetikzlibrary{arrows,positioning} 
\tikzset{
    >=stealth',
    punkt/.style={
           rectangle,
           rounded corners,
           draw=black, very thick,
           text width=4.5em,
           minimum height=2em,
           text centered},
    another/.style={
      rectangle,
      draw=black, very thick,
      text width=6.5em,
      minimum height=2em,
      text centered},
    pil/.style={
           ->,
           thick,
           shorten <=2pt,
           shorten >=2pt,},
    dotted/.style={
      ->,
      thick,
      dashed,
      shorten <=2pt,
      shorten >=2pt,}
}

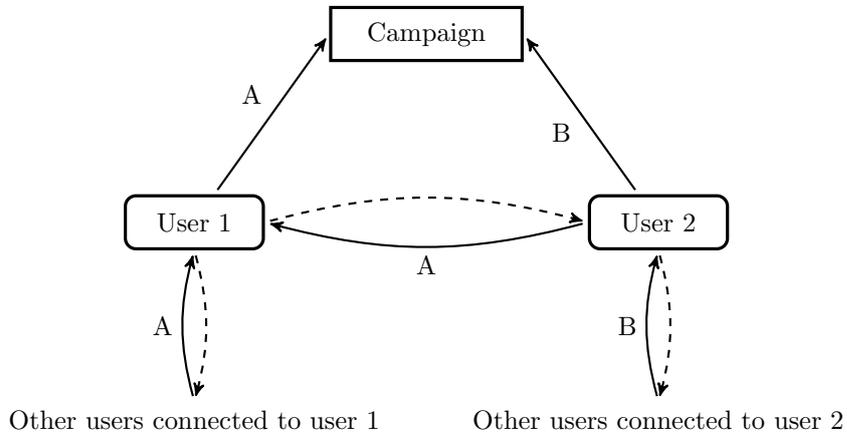
\begin{figure}[htb!]
\begin{center}
\begin{tikzpicture}[node distance=2cm, auto, ]
  \node[another] (campaign) {Campaign};
  \node[below=of campaign] (dummy0) {};
  \node[punkt, left=of dummy0] (user1) {User 1};
  \node[below=of user1] (dummy1) {Other users connected to user 1};
  \node[punkt, right=of dummy0] (user2) {User 2};
  \node[below=of user2] (dummy2) {Other users connected to user 2};
  \path[pil] (user1) edge node {A} (campaign.west);
  \path[pil] (user2) edge node {B} (campaign.east);
  \path[pil, bend left=15] (user2.west) edge node {A} (user1.east);
  \path[dotted, bend left=15] (user1.east) edge node {} (user2.west);
  \path[pil, bend left=15] (dummy1.north) edge node {A} (user1.south);
  \path[dotted, bend left=15] (user1.south) edge node {} (dummy1.north);
  \path[pil, bend left=15] (dummy2.north) edge node {B} (user2.south);
  \path[dotted, bend left=15] (user2.south) edge node {} (dummy2.north);
\end{tikzpicture}
\end{center}
\caption{Flow diagram of App Movement sharing. A straight arrow represents the generation of the labelled URL when the user visits the campaign page, while each solid-dashed curved arrow pair represents the click of the labelled URL and the direction of infection, respectively.} \label{fig.flow}
\end{figure}

\begin{knitrout}
\definecolor{shadecolor}{rgb}{0.969, 0.969, 0.969}\color{fgcolor}\begin{figure}

{\centering \includegraphics[width=\maxwidth]{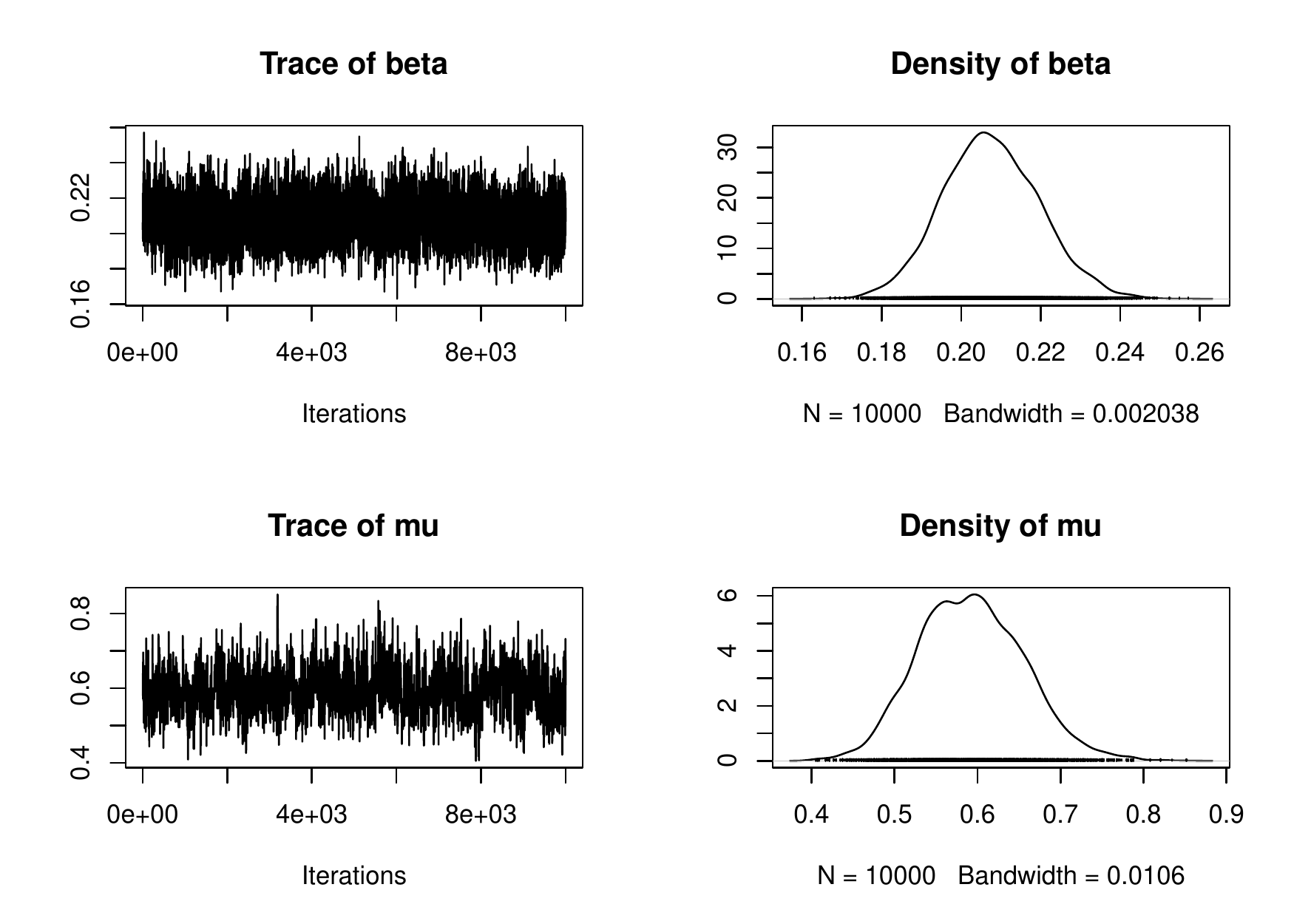} 

}

\caption{Traceplots and posterior densities of $\beta$ and $\mu$ of the PA model\newline{for} the epidemic shown in Figures \ref{fig.eda_I} and \ref{fig.eda_P}.}\label{fig.mcmc}
\end{figure}

\end{knitrout}
\clearpage

The model is fitted to each of ten campaign epidemics separately, assuming they have no influence on each other for simplicity, with $m$ ranging from 334 to 402. Each campaign corresponds to a different proposed application. The inference algorithm is used with $\gamma$ fixed to 0. For each epidemic, 5 chains of length 2000 (no thinning) are obtained, after the first 1000 iterations being discarded as burn-in, during which the proposal standard deviation for $\mu$ is tuned. The traceplots and posterior densities of $\beta$ and $\mu$ are plotted in Figure \ref{fig.mcmc}, for the model fit to the epidemic visualised in Figures \ref{fig.eda_I} and \ref{fig.eda_P}. The acceptance rate for $\mu$ is 0.269, and is similar for the other 9 epidemics considered. The posterior means and standard deviations of $\beta$, $\mu$ and $\alpha$ for all epidemics are reported in Table \ref{tab:summary_pat}. Also reported is the correlation between $\beta$ and $\mu^{*}=\mu+e^{-\mu}$, which is modest but consistently negative. For any parameter $\theta$, we denote E$(\theta|\mathcal{P},\tilde{\mathbf{I}})$ as its posterior mean. We can see that E$(\alpha|\mathcal{P},\tilde{\mathbf{I}})$ is not dependent on $m$ and is significantly different across the epidemics. Combining with the fact that the correlation with $\mu$ (or $\mu^{*}$) is modest (not shown), $\alpha$ can be seen to be successfully identified. 

\begin{center}
\begin{knitrout}
\definecolor{shadecolor}{rgb}{0.969, 0.969, 0.969}\color{fgcolor}\begin{table}

\caption{\label{tab:summary_pat}Posterior mean (standard deviation) and correlation of the scalar parameters in the PA model fitted to ten different campaign epidemics.}
\centering
\begin{tabular}[t]{|c|c||l|l|l||l||c|c||l|l|l||l||c|c||l|l|l||l||c|c||l|l|l||l||c|c||l|l|l||l||c|c||l|l|l||l}
\hline
epidemic & m & beta & mu & correlation & alpha\\
\hline
1 & 402 & 0.091 (0.005) & 0.323 (0.051) & -0.041 & 0.096 (0.005)\\
\hline
2 & 391 & 0.218 (0.012) & 0.775 (0.069) & -0.137 & 0.269 (0.016)\\
\hline
3 & 390 & 0.384 (0.022) & 0.606 (0.057) & -0.147 & 0.442 (0.025)\\
\hline
4 & 388 & 0.242 (0.013) & 0.689 (0.061) & -0.136 & 0.289 (0.017)\\
\hline
5 & 387 & 0.315 (0.017) & 0.718 (0.061) & -0.071 & 0.38 (0.022)\\
\hline
6 & 371 & 0.491 (0.028) & 0.59 (0.058) & -0.087 & 0.563 (0.033)\\
\hline
7 & 363 & 0.373 (0.022) & 0.603 (0.061) & -0.115 & 0.43 (0.026)\\
\hline
8 & 358 & 0.453 (0.026) & 0.708 (0.061) & -0.068 & 0.545 (0.033)\\
\hline
9 & 335 & 0.208 (0.012) & 0.592 (0.063) & -0.100 & 0.238 (0.015)\\
\hline
10 & 334 & 0.147 (0.009) & 0.601 (0.066) & -0.169 & 0.169 (0.01)\\
\hline
\end{tabular}
\end{table}

\end{knitrout}
\end{center}
\begin{center}
\begin{table}

\caption{\label{tab:summary_brg}Posterior mean (standard deviation) and correlation of the parameters in the BRG model fitted to ten different campaign epidemics. }
\centering
\begin{tabular}[t]{|c|c||l|l|l||c|c||l|l|l||c|c||l|l|l||c|c||l|l|l||c|c||l|l|l}
\hline
epidemic & m & beta & p & correlation\\
\hline
1 & 402 & 0.087 (0.005) & 0.0059 (0.00029) & -0.085\\
\hline
2 & 391 & 0.232 (0.012) & 0.006 (0.00031) & -0.069\\
\hline
3 & 390 & 0.411 (0.022) & 0.0055 (0.00028) & -0.052\\
\hline
4 & 388 & 0.262 (0.014) & 0.0058 (0.00029) & -0.045\\
\hline
5 & 387 & 0.338 (0.018) & 0.0057 (0.00029) & -0.029\\
\hline
6 & 371 & 0.512 (0.028) & 0.0058 (0.0003) & -0.036\\
\hline
7 & 363 & 0.399 (0.022) & 0.0059 (0.00032) & -0.069\\
\hline
8 & 358 & 0.492 (0.027) & 0.0061 (0.00032) & -0.033\\
\hline
9 & 335 & 0.219 (0.013) & 0.0065 (0.00036) & -0.052\\
\hline
10 & 334 & 0.154 (0.009) & 0.0066 (0.00037) & -0.106\\
\hline
\end{tabular}
\end{table}

\vspace{-3cm}
\end{center}

Model comparison or selection is difficult here because the BRG model by {\cite{bo02}} is not nested in our proposed model, even when $\gamma$ is treated as a parameter. Nevertheless, we fit the BRG model to the same campaign epidemics, focusing on the parameter results to examine its goodness-of-fit. The posterior means and standard deviations of the parameters are reported in Table \ref{tab:summary_brg}, which shows that E$(p|\mathcal{P},\tilde{\mathbf{I}})$ is of the same magnitude across all the epidemics. For the singled out epidemic with $m=$ 335, compared to the average degree E$(p|\mathcal{P},\tilde{\mathbf{I}})\times(m-1)=$ 2.158, which means each user on average is connected to slightly more than two other users, the most infectious user has infected 18 other users. If we use E$(p|\mathcal{P},\tilde{\mathbf{I}})$ as the true value of $p$ and simulate a BRG, the probability that one particular user is connected to at least 18 users is \ensuremath{1.453\times 10^{-11}}. Combining these two quantities with the independence of potential edges, we can see that it is very unlikely a BRG generated in this way will be connected, let alone overlay $\mathcal{P}$. On the other hand, the network construction described in Section \ref{sect.model} ensures that the PA network generated is always connected. Finally, contrary to the clear inverse relationship between $\beta$ and $p$ reported in {\cite{bo02}} for both simulated and real-life data, the joint posterior of $(\beta,p)$ can be well approximated by a bivariate Gaussian distribution, for all epidemics reported here. Combining with the fact that the correlations are small (last column of Table \ref{tab:summary_brg}), $\beta$ and $p$ can be said to be close to independence \textit{aposteriori}. This suggests that the presence of $\mathcal{P}$ actually makes $p$ (and $\beta$) identifiable, but the estimate of the successfully identified $p$ now shows a poor fit of the BRG model to our data.

While the values of $E(p|\mathcal{P},\tilde{\mathbf{I}})$ in Table \ref{tab:summary_brg} are low, those of E$(\beta|\mathcal{P},\tilde{\mathbf{I}})$ are similar to their PA counterparts in Table \ref{tab:summary_pat}, but are unusually high compared to real-life epidemics. This is because, while real-life epidemics usually spanned days (see, for example, {\cite{bo02}} and {\cite{nr05}}), the campaign epidemics spanned weeks (see the time scale of Figure \ref{fig.eda_I}). Out of the ten epidemics reported here, epidemics 1 and 6 spanned the longest and shortest, with a period of 187.371 and 36.7 days, respectively, and this explains why their respective values of $E(\beta|\mathcal{P},\tilde{\mathbf{I}})$ are on opposite extremities among those reported in Table \ref{tab:summary_pat}.

Using the posterior of $\beta$ and $\mu$, we can simulate the network epidemic and obtain the predictive distribution of the cumulative counts over time, of which the 95\% predictive intervals (PI) overlay the observed data in Figure \ref{fig.app_I}. While the early period of the epidemic lies within the 95\% PI, it is the slower periods of infections towards the end of the observation period that are more difficult to reproduce.

\begin{knitrout}
\definecolor{shadecolor}{rgb}{0.969, 0.969, 0.969}\color{fgcolor}\begin{figure}[htb!]

{\centering \includegraphics[width=\maxwidth]{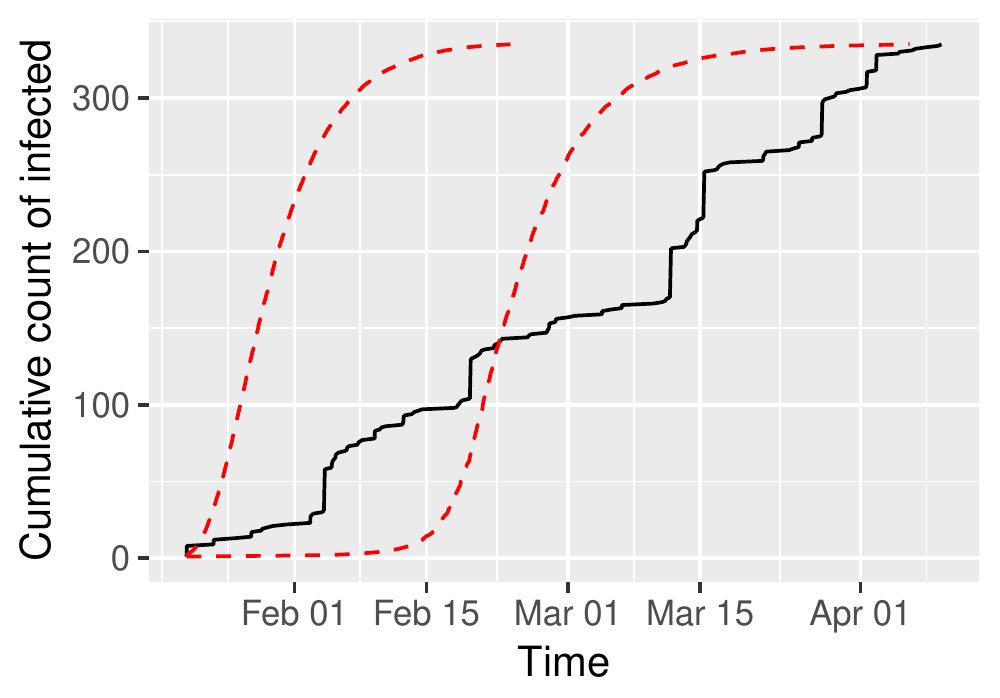} 

}

\caption{Cumulative count of infected for the campaign epidemic shown in Figure \ref{fig.eda_I}, overlaid by 95\% predictive intervals of simulated cumulative counts over time (red dashed lines).}\label{fig.app_I}
\end{figure}

\end{knitrout}

\section{Discussion} \label{sect.con}
We have described a network epidemic model which combines the SI epidemic model with the PA network model, with the inference carried out by MCMC as the likelihood can be explicitly computed. The results of two simulation studies suggest dropping\deleted{ of} one parameter in order to make the \textit{network scaled epidemic rate} parameter identifiable. The model and inference algorithm are successfully applied to ten different ``sharing'' epidemics of a set of online community commissioning data. The results suggest that the PA model is a better alternative for network generation than the BRG model, for data sets of epidemics taking place on social networks.

Several modifications can potentially make the model more useful. First of all, information on the average connectedness or the degree distribution on social networks can be solicited beforehand, so that an informative prior can be assigned to $\mu$ and/or $\gamma$, at least one of which can then be identified. Given the vast amount of data about social networks such as Facebook and Twitter freely available on the Internet, such information should be possible to obtain.

Another way of gaining information for the parameters, particularly for the App Movement data, is combining the epidemics in the modelling and inference. As users on the social network are usually involved in more than one epidemic, we can pull together several epidemics which have overlap in the users, and build a larger underlying network $\mathcal{G}$ comprising all the users involved, which is guaranteed to be connected. As a user may be infected in one epidemic but not another, each of the epidemics may then be incomplete. While the likelihood calculations, for example \eqref{eqn.lik_i}, may differ slightly, and each epidemic has a different rate parameter (no matter whether it is network scaled or not), the inference procedure is basically the same, and only one set of parameters $(\mu, \gamma)$ is used to govern the network generation. Borrowing strength from other epidemics in this way will utilise more information available in the data, and result in parameters being better identified or more precisely estimated.

That the epidemic is Markovian given the network is a simplistic assumption, which has been shown inadequate for real-life epidemics. To relax this assumption, one can use alternative distributions for the infectious period, such as the Gamma distribution, which is used by, for example, \cite{xn14}. This is not relevant here because the SI model is used instead of the more popular SIR model, or any compartment model in which being infectious is not the final state. Instead of using one single epidemic rate $\beta$ for all the infections, meaning that the interarrival times are exponentially distributed, one can use a different rate $\beta_i$ for the infection of individual $i$, where $\beta_i$ is drawn from certain probability distribution. This approach to modelling non-Markovian processes, proposed by \cite{mr17}, can be applied to any kind of compartment model, and can encompass a range of interarrival time distributions, simply by choosing a different probability distribution from which $\beta_i$ is drawn. When it comes to the inference, the interarrival time distribution parameter(s) and each $\beta_i$, given all other epidemic rate parameters, will be updated individually, on top of the existing parameters and latent variables. Given that most of the computational time lies in updating the potential edges of $\mathcal{G}$ one by one, this should not add much to the computational burden.\added{ On the other hand, including users who have not seen or joined the campaign to correct for the potential overestimation of $\beta$ will add to the computation burden, because the number of users \textit{not} infected by a particular campaign is vast compared to the number of infected.}


\section*{Acknowledgement}
This research was funded by the EPSRC grant DERC: Digital Economy Research Centre (EP/M023001/1). 

\bibliographystyle{agsm}
\bibliography{ref_am}

\appendix

\section{Likelihood derivation} \label{sect.appendix_lik}
This appendix derives each componet of the likelihood, in reverse order of \eqref{eqn.lik}.

First, the quantity $\pi(\mathcal{G}|\mu,\gamma,\boldsymbol{\sigma})$ can be divided into two components, which correspond to the processes described in Sections \ref{sect.model_edges} and \ref{sect.model_network}, respectively. To derive the contribution of the independent sequence of random numbers of new edges to $\pi(\mathcal{G}|\mu,\gamma,\boldsymbol{\sigma})$, we first establish its relationship with $\mathcal{G}$. Specifically, for $3\leq i\leq m$, $X_i$ is the sum of the first $i-1$ elements of the $i^{\text{th}}$ row (or column) of $\mathcal{G}$, that is, $X_i=\sum_{j=1}^{i-1}\mathcal{G}_{\sigma_i\sigma_j}$. As $\mathcal{G}_{\sigma_1\sigma_2}=1$, we have
\begin{align}
  \sum_{i=3}^mX_i&=\sum_{i=3}^m\sum_{j=1}^{i-1}\mathcal{G}_{\sigma_i\sigma_j}
  =\sum_{i=2}^m\sum_{j=1}^{i-1}\mathcal{G}_{\sigma_i\sigma_j}-1.\nonumber
\end{align}
This makes sense as the sum of new edges is equal to the total number of edges minus one. Using \eqref{eqn.edges}, we can calculate the likelihood of the sequence of random numbers of new edges:
\begin{align}
  L_1(\mathcal{G};\boldsymbol{\sigma},\mu):=~&\mathlarger{\mathlarger\prod_{i=3}^m}
  \left\{
  \left[e^{-\mu}(1+\mu)\right]^{\boldsymbol{1}\{X_i=1\}}
  \left(\frac{e^{-\mu}\mu^{X_i}}{X_i!}\right)^{\boldsymbol{1}\{2\leq X_i<i-1\}}
  \left(\sum_{z=i-1}^\infty\frac{e^{-\mu}\mu^z}{z!}\right)^{\boldsymbol{1}\{X_i=i-1\}}
  \right\}\nonumber \\
  =~&e^{-(m-2)\mu}\times\mathlarger{\mathlarger\prod_{i=3}^{m}}
  \left\{
  \frac{\mu^{X_i}}{X_i!}\left(\frac{1+\mu}{\mu}\right)^{\boldsymbol{1}\{X_i=1\}}
  \left[\left(\sum_{z=i-1}^{\infty}\frac{\mu^z}{z!}\right)\left/\frac{\mu^{i-1}}{(i-1)!}\right.\right]^{\boldsymbol{1}\{X_i=i-1\}}
  \right\}\nonumber \\
  =~& \left[\mathlarger\prod_{i=3}^m\left(\mathlarger\sum_{j=1}^{i-1}\mathcal{G}_{\sigma_i\sigma_j}\right)!\right]^{-1} \times
  \mathlarger{e^{-(m-2)\mu}\mu^{|\mathcal{G}|-1}}
  \left(\frac{1+\mu}{\mu}\right)^{\sum_{i=3}^m\boldsymbol{1}\{\sum_{j=1}^{i-1}\mathcal{G}_{\sigma_i\sigma_j}=1\}}
  \nonumber\\*
  &\times \mathlarger\prod_{i=3}^m \left[\left(\sum_{z=i-1}^{\infty}\frac{\mu^z}{z!}\right)\left/\frac{\mu^{i-1}}{(i-1)\mathlarger!}\right.\right]^{\boldsymbol{1}\{\sum_{j=1}^{i-1}\mathcal{G}_{\sigma_i\sigma_j}=i-1\}}\label{eqn.appendix_lik_L1}
\end{align}
where $\boldsymbol{1}\{A\}$ is the indicator function of event $A$, and $|\mathcal{G}|:=\sum_{i=2}^m\sum_{j=1}^{i-1}\mathcal{G}_{\sigma_i\sigma_j}$.

For the process of attaching edges to nodes, it is straightforward to compute the likelihood using \eqref{eqn.model_weights}. However, because of the nature of weighted sampling without replacement, we have to, for each $i$, calculate the probability conditional on each of the $X_i!$ permutations of the selected nodes and then average over all $X_i!$ probabilities to arrive at the likelihood. As calculating the exact likelihood in this way is not computationally feasible because the factorial grows faster than the exponential function, we approximate the likelihood based on one permutation of weighted sampling without replacement instead. The contribution by the new edges brought by node $i$ is
\begin{align}
  L_{2i}&=X_i!\times w_1^{\mathcal{G}_{\sigma_i\sigma_1}}\mathlarger\prod_{j=2}^{i-1}\left(\frac{w_j}{1-\sum_{k=1}^{j-1}w_k}\right)^{\mathcal{G}_{\sigma_i\sigma_j}} \nonumber\\
  &=\left(\sum_{j=1}^{i-1}\mathcal{G}_{\sigma_i\sigma_j}\right)\mathlarger!\times
  w_1^{\mathcal{G}_{\sigma_i\sigma_1}}\mathlarger\prod_{j=2}^{i-1}\left(\frac{w_j}{1-\sum_{k=1}^{j-1}w_k}\right)^{\mathcal{G}_{\sigma_i\sigma_j}}, \nonumber
\end{align}
where $w_j$ is given by \eqref{eqn.model_weights}. Therefore, the likelihood of the process of adding new edges is
\begin{align}
  L_2(\mathcal{G};\boldsymbol{\sigma},\gamma):=&\mathlarger\prod_{i=3}^mL_{2i}\nonumber \\
  =&\mathlarger\prod_{i=3}^m\left(\sum_{j=1}^{i-1}\mathcal{G}_{\sigma_i\sigma_j}\right)\mathlarger! \times
  \mathlarger\prod_{i=3}^m\left[w_1^{\mathcal{G}_{\sigma_i\sigma_1}}\mathlarger\prod_{j=2}^{i-1}\left(\frac{w_j}{1-\sum_{k=1}^{j-1}w_k}\right)^{\mathcal{G}_{\sigma_i\sigma_j}}\right].\label{eqn.appendix_lik_L2}
\end{align}
Multiplying \eqref{eqn.appendix_lik_L1} and \eqref{eqn.appendix_lik_L2} gives the expression of $\pi(\mathcal{G}|\mu,\gamma,\boldsymbol{\sigma})$ in \eqref{eqn.lik_L1L2} as $\mathlarger\prod_{i=3}^m\left(\mathlarger\sum_{j=1}^{i-1}\mathcal{G}_{\sigma_i\sigma_j}\right)\mathlarger!$ cancels.

Second, $\pi(\tilde{\mathbf{I}}|\mathcal{G},\beta)$ contains contributions from the $m-1$ infections and from the connections through which no infections occurred:
\begin{align}
  \pi(\tilde{\mathbf{I}}|\mathcal{G},\beta)&=\mathlarger\prod_{i=1}^{m-1}\mathlarger\prod_{j=i+1}^{m}\left(\beta\exp[-\beta(\tilde{I}_j-\tilde{I}_i)]\right)^{\boldsymbol{1}\{\mathcal{G}_{ij}=1,\mathcal{P}_{ij}=1\}}\left(\exp[-\beta(\tilde{I}_j-\tilde{I}_i)]\right)^{\boldsymbol{1}\{\mathcal{G}_{ij}=1,\mathcal{P}_{ij}\neq1\}}\nonumber\\
  &=\beta^{m-1}\exp\left(-\beta\sum_{i=1}^{m-1}\sum_{j=i+1}^{m}\left[(\tilde{I}_j-\tilde{I}_i)\boldsymbol{1}\{\mathcal{G}_{ij}=1\}\right]\right) \nonumber \\
  &=\beta^{m-1}\exp\left(-\beta\sum_{i=1}^{m-1}\sum_{j=i+1}^{m}\mathcal{G}_{ij}(\tilde{I}_j-\tilde{I}_i)\right), \nonumber
\end{align}
which is the same as \eqref{eqn.lik_i}, and confirms that the infection times are independent of the transmission tree, as the infection mechanism is a Poisson process \citep{bo02}, hence the dropping of $\mathcal{P}$ in the function argument.

Third, the contribution to likelihood by $\mathcal{P}$ is straightforward, as $\pi(\mathcal{P}|\mathcal{G})$ ``is the uniform distribution on the set of all possible infection pathways'' \citep{bo02}. We have to ensure nobody infects the initial infected (individual 1), and for each individual $j$ of the remaining $m-1$ individuals, there is one and only one neighbour who is infected prior to individual $j$, ends up infecting individual $j$. Therefore,
\begin{align}
\pi(\mathcal{P}|\mathcal{G}) &= \mathbf{1}\{\text{nobody infects individual 1}\} \nonumber \\*
&\times\mathlarger\prod_{j=2}^m\frac{\mathbf{1}\{\text{only one previously infected neighbour infects individual $j$}\}}{\text{number of neighbours infected before individual $j$}} \nonumber \\
&=\mathlarger\prod_{i=2}^m\mathbf{1}\{\mathcal{P}_{i1}=0\}\times
\mathlarger\prod_{j=2}^m \frac{\mathbf{1}\left\{\displaystyle\sum_{i=1}^{j-1}\mathcal{P}_{ij}=1\right\}\displaystyle\prod_{i=j}^m\mathbf{1}\left\{\mathcal{P}_{ij}=0\right\} \prod_{i=1}^m\mathbf{1}\left\{\mathcal{P}_{ij}\leq\mathcal{G}_{ij}\right\}}
                  {\displaystyle\sum_{i=1}^{j-1}\mathcal{G}_{ij}} \nonumber \\
&=\mathlarger\prod_{j=1}^{m-1}\mathlarger\prod_{i=j+1}^{m}(1-\mathcal{P}_{ij})\times
\mathlarger\prod_{j=2}^{m}\frac{\mathbf{1}\left\{\displaystyle\sum_{i=1}^{j-1}\mathcal{P}_{ij}=1\right\}}{\displaystyle\sum_{i=1}^{j-1}\mathcal{G}_{ij}}\times
\mathlarger\prod_{j=2}^{m}\mathlarger\prod_{i=1}^{j-1}\mathbf{1}\left\{\mathcal{P}_{ij}\leq\mathcal{G}_{ij}\right\}, \nonumber
\end{align}
which is the same as \eqref{eqn.lik_p}.

\section{MCMC algorithm} \label{sect.appendix_mcmc}
This appendix describes the MCMC algorithm for the inference outlined in Section \ref{sect.lik_inf}.

\textbf{Sampling $\beta$}: As we have assigned a (conditional) conjugate prior to $\beta$, its full conditional posterior is given by
\begin{align}
\beta|\cdots\sim\text{Gamma}\left(a_{\beta}+m-1,~b_{\beta}+\mathlarger\sum_{1\leq i<j\leq m}\mathcal{G}_{ij}(\tilde{I}_j-\tilde{I}_i)\right),\nonumber
\end{align}
where $\cdots$ means all arguments in \eqref{eqn.inf_joint} other than the quantity of interest. So, conditional upon the data and all other parameters, $\beta$ can be sampled via a Gibbs step.

\textbf{Sampling $\mu$}: As $\mu$ is only involved in one single term in the likelihood, namely $L_1(\mathcal{G};\boldsymbol{\sigma},\mu)$, in the absence of a conjugate prior, we can update $\mu$ using a simple Metropolis step. Specifically, we propose $\mu^{*}$ from a symmetrical proposal $q(\cdot|\mu)$, and accept $\mu^{*}$ with probability given by
\begin{align}
  \alpha_{\mu}&=1\wedge\frac
        {L_1(\mathcal{G};\boldsymbol{\sigma},\mu^{*})\times(\mu^{*})^{a_\mu-1}\exp(-b_\mu\mu^{*})\boldsymbol{1}\{\mu^{*}>0\}}
        {L_1(\mathcal{G};\boldsymbol{\sigma},\mu)\times(\mu)^{a_\mu-1}\exp(-b_\mu\mu)\boldsymbol{1}\{\mu>0\}}.\nonumber
\end{align}
\textbf{Sampling $\gamma$}: The Metropolis step for $\gamma$ is similar to that for $\mu$, as the former is involved in $L_2(\mathcal{G};\boldsymbol{\sigma},\gamma)$ in the likelihood. We propose $\gamma^{*}$ from a symmetrical proposal $q(\cdot|\gamma)$ and accept $\gamma^{*}$ with probability
\begin{align}
  \alpha_{\gamma}&=1\wedge\frac
        {L_2(\mathcal{G};\boldsymbol{\sigma},\gamma^{*})\times\boldsymbol{1}\{0\leq\gamma^{*}\leq1\}}
        {L_2(\mathcal{G};\boldsymbol{\sigma},\gamma)\times\boldsymbol{1}\{0\leq\gamma\leq1\}}.\nonumber
\end{align}
\textbf{Sampling $\boldsymbol{\sigma}$}: To update the ordering as a whole, we propose $\boldsymbol{\sigma}^{*}$, which is accepted with probability
\begin{align}
  \alpha_{\boldsymbol{\sigma}}&=1\wedge\frac
        {\pi(\mathcal{G}|\mu,\gamma,\boldsymbol{\sigma}^{*})}
        {\pi(\mathcal{G}|\mu,\gamma,\boldsymbol{\sigma})}.\nonumber
\end{align}
This requires a symmetrical proposal on the permutation space. Specifically, we use a ``random walk by insertion'' method used by {\cite*{bks06}}. Two indices $i$ and $j$ are first sampled with replacement from $\{1,2,\ldots,m\}$ uniformly. Without loss of generality, assume that $i<j$. While the current ordering is
\begin{align}
  \boldsymbol{\sigma}=(\sigma_1,\ldots,\sigma_{i-1},\sigma_{i},\sigma_{i+1},\ldots,\sigma_{j-1},\sigma_{j},\sigma_{j+1},\ldots,\sigma_m),\nonumber
\end{align}
the proposed ordering is
\begin{align}
  \boldsymbol{\sigma}^{*}=(\sigma_1,\ldots,\sigma_{i-1},\sigma_{i+1},\ldots,\sigma_{j-1},\sigma_{j},\sigma_{i},\sigma_{j+1},\ldots,\sigma_m).\nonumber
\end{align}
The intuition is that the $i^{\text{th}}$ card of a deck of cards is taken out and inserted in the $j^{\text{th}}$ position. As $(i,j)$ and $(j,i)$ have the same probability of being sampled in their particular orders, the proposal is symmetrical. This method is, according to {\cite{bks06}}, more efficient than the random swap method, in which an arbitrary pair of adjacenct indices $(\sigma_i,\sigma_{i+1})~(1\leq i<m)$ is picked, and a swap between them produces the proposed ordering.

Theoretical properties are not clear yet to provide guildlines on optimising the number of random insertions in each MCMC iteration. As it is found out that the majority of the computation time per iteration is taken by updating all potential edges of $\mathcal{G}$ individually, which will be described below, we simply propose to update the ordering $m$ times in each iteration, so that each index will on average be picked and inserted once. It should however be noted that an index potentially changes its position even if it is not selected, as long as its position lies between $i$ and $j$ inclusive.

\textbf{Sampling $\mathcal{G}$}: We will use a Gibbs step to update each of the $\binom m 2$ potential edges in $\mathcal{G}$ sequentially, and this requires defining the quantities required first. Unlike a BRG in \cite{oneill02}, \cite{nr05}, \cite{rm08} and \cite{gwh11}, the potential edges of $\mathcal{G}$ are not independent anymore, both \textit{apriori} and \textit{aposteriori}. Still, we can update each potential edge $\mathcal{G}_{ij}~(1\leq i<j\leq m)$, conditional on all of $\mathcal{G}$ except $\mathcal{G}_{ij}$ (and $\mathcal{G}_{ji}$ because of symmetry), denoted by $\mathcal{G}_{-ij}$. While $\mathcal{G}_{-ij}$ is not a proper adjacency matrix, we also define matrices $\mathcal{G}_{-ij}^0:=\{\mathcal{G}_{-ij},\mathcal{G}_{ij}=0\}$ and $\mathcal{G}_{-ij}^1:=\{\mathcal{G}_{-ij},\mathcal{G}_{ij}=1\}$, so that exactly one of $\mathcal{G}_{-ij}^0$ and $\mathcal{G}_{-ij}^1$ is identical to $\mathcal{G}$.

Because of the difference in the network ordering and epidemic ordering, for each pair $(i,j)$, we proceed to sample $\mathcal{G}_{\sigma_i\sigma_j}$ instead of $\mathcal{G}_{ij}$. This will not pose a problem in practice as we will go through all combinations of $(i,j)$ satisfying $1\leq i<j\leq m$. Because of the 1-1 relationship between $(i,j)$ and $(\sigma_i,\sigma_j)$ given $\boldsymbol{\sigma}$, eventually all the potential edges will be updated. For notational convenience, we also define $s=\min(\sigma_i,\sigma_j)$ and $t=\max(\sigma_i,\sigma_j)$, which implies $\tilde{I}_t>\tilde{I}_s$. If $\mathcal{P}_{st}=1$, as mentioned in Section \ref{sect.model_epidemic} and implied by \eqref{eqn.lik_p}, the four equivalent quantities, namely $\mathcal{G}_{\sigma_i\sigma_j},\mathcal{G}_{\sigma_j\sigma_i},\mathcal{G}_{st}$ and $\mathcal{G}_{ts}$, are equal to 1 with posterior probability 1, regardless of all other parameters and $\mathcal{G}_{-st}$. Therefore, we shall only consider $\pi(\mathcal{G}_{st}|\mathcal{P}_{st}=0,\mathcal{G}_{-st},\cdots)$ in detail. Before doing so, we observe that $\beta$ can be integrated out in the joint posterior in \eqref{eqn.inf_joint}, which is achieved by substituting \eqref{eqn.lik_i} and the prior of $\beta$ in \eqref{eqn.inf_prior} into \eqref{eqn.inf_post}, followed by integration with respect to $\beta$:
\begin{align}
\pi(\mathcal{G},\beta,\mu,\gamma,\boldsymbol{\sigma}|\mathcal{P},\tilde{\mathbf{I}})
\propto&~\pi(\mathcal{P}|\mathcal{G})\times 
\beta^{m-1}\exp\left(-\beta\mathlarger\sum_{1\leq i<j\leq m}\mathcal{G}_{ij}(\tilde{I}_j-\tilde{I}_i)\right)
\times\pi(\mathcal{G}|\mu,\gamma,\boldsymbol{\sigma}) \nonumber \\*
&\times\beta^{a_\beta-1}\exp(-\beta b_\beta)\times\pi(\mu)\pi(\gamma)\pi(\boldsymbol{\sigma}) \nonumber\\
\int\pi(\mathcal{G},\beta,\mu,\gamma,\boldsymbol{\sigma}|\mathcal{P},\tilde{\mathbf{I}})d\beta
\propto&~\pi(\mathcal{P}|\mathcal{G})\times\pi(\mathcal{G}|\mu,\gamma,\boldsymbol{\sigma})\times
\pi(\mu)\pi(\gamma)\pi(\boldsymbol{\sigma})\nonumber\\*
&\times\int\beta^{(a_\beta+m-1)-1}\exp\left[-\beta\left(b_\beta+\mathlarger\sum_{1\leq i<j\leq m}\mathcal{G}_{ij}(\tilde{I}_j-\tilde{I}_i)\right)\right]d\beta\nonumber\\
\pi(\mathcal{G},\mu,\gamma,\boldsymbol{\sigma}|\mathcal{P},\tilde{\mathbf{I}})d\beta
\propto&~\pi(\mathcal{P}|\mathcal{G})\times\pi(\mathcal{G}|\mu,\gamma,\boldsymbol{\sigma})\times
\pi(\mu)\pi(\gamma)\pi(\boldsymbol{\sigma})\nonumber\\*
&\times\left(b_\beta+\mathlarger\sum_{1\leq i<j\leq m}\mathcal{G}_{ij}(\tilde{I}_j-\tilde{I}_i)\right)^{-(a_\beta+m-1)}.\label{eqn.int_beta_out}
\end{align}
The last line is the reciprocal of the constant of proportionality as the integrand in the second line is the density function of a Gamma$\left(a_\beta+m-1,b_\beta+\mathlarger\sum_{1\leq i<j\leq m}\mathcal{G}_{ij}(\tilde{I}_j-\tilde{I}_i)\right)$ distribution without the constant. With $\beta$ integrated out $\pi(\mathcal{G}_{st}|\mathcal{P}_{st}=0,\mathcal{G}_{-st},\cdots)$ can now be derived. Using \eqref{eqn.lik_p}, \eqref{eqn.appendix_lik_L1} and \eqref{eqn.appendix_lik_L2}, we have
\begin{align}
  \Pr(\mathcal{G}_{st}=0|\mathcal{P}_{st}=0,\mathcal{G}_{-st},\cdots)&\propto\frac
     {\pi(\mathcal{G}_{-st}^0|\mu,\gamma,\boldsymbol{\sigma})}
     {\displaystyle\sum_{k=1,k\neq s}^{t-1}\mathcal{G}_{kt}} 
     \left(b_\beta+\mathlarger\sum_{\substack{1\leq i<j\leq m\\i\neq s,j\neq t}}\mathcal{G}_{ij}(\tilde{I}_j-\tilde{I}_i)\right)^{-(a_\beta+m-1)},\nonumber\\
  \Pr(\mathcal{G}_{st}=1|\mathcal{P}_{st}=0,\mathcal{G}_{-st},\cdots)&\propto\frac
     {\pi(\mathcal{G}_{-st}^1|\mu,\gamma,\boldsymbol{\sigma})}
     {\displaystyle\sum_{k=1,k\neq s}^{t-1}\mathcal{G}_{kt}+1}
     \left(b_\beta+\mathlarger\sum_{\substack{1\leq i<j\leq m\\i\neq s,j\neq t}}\mathcal{G}_{ij}(\tilde{I}_j-\tilde{I}_i) + (\tilde{I}_t-\tilde{I}_s)\right)^{-(a_\beta+m-1)}.\nonumber
\end{align}
The posterior distribution $\pi(\mathcal{G}_{st}|\mathcal{P}_{st}=0,\mathcal{G}_{-st},\cdots)$ can then be obtained directly. The interpretation of the right-hand expression in the two lines is as follows. The numerator and the denominator correspond to the network likelihood and the transmission tree likelihood, respectively, under the corresponding scenario of whether nodes $\sigma_i$ and $\sigma_j$ are network neighbours, which also affects the summation in \eqref{eqn.int_beta_out}, hence the difference between the two lines in the bracketed term. It can be seen that the computational burden lies in computing the network likelihood. Assume that we are going to update $\mathcal{G}_{st}$, whose current value is $x\in\{0,1\}$, and that we have retained the likelihood under $\mathcal{G}=\mathcal{G}_{-st}^x$ after updating the previous potential edge. We now have to calculate the network likelihood under the alternative scenario $\mathcal{G}=\mathcal{G}_{-st}^{1-x}$. The summations involved in \eqref{eqn.appendix_lik_L1} and \eqref{eqn.appendix_lik_L2} makes it not possible to factorise the network likelihood, which therefore requires the whole of $\mathcal{G}_{-st}^{1-x}$ to compute.

\textbf{Sampling $\mathcal{P}$}: Although the inference algorithm above is for the transmission tree being part of the data, it can be extended to include sampling of $\mathcal{P}$ if it is unknown, in a way similar to how $\mathcal{G}$ is being treated as latent variables and inferred. As mentioned in Section \ref{sect.lik_inf}, $\mathcal{P}$ follows a uniform distribution on all possible infection pathways given $\mathcal{G}$, thus independent of how $\mathcal{G}$ is generated in the first place. As the same Gibbs step for sampling $\mathcal{P}$ described by {\cite{bo02}} can therefore be used, it will not be repeated here.

\end{document}